\documentclass[openany]{nature}
\usepackage{caption}
\usepackage{graphicx}
\usepackage[caption = false]{subfig}
\usepackage{academicons}
\usepackage{amsmath, amssymb, url, xcolor}
\usepackage{longtable}
\usepackage{CJKutf8}
\usepackage{adjustbox}
\usepackage{hyperref}
\usepackage{threeparttable}
\usepackage{ulem}
\usepackage{rotating}
\usepackage{booktabs}
\usepackage{multirow}
\usepackage{makecell}

\usepackage{caption}

\usepackage{tabularx}
\usepackage{array}

\usepackage[nolist]{acronym}
\graphicspath{{./}{figures/}}
\usepackage{lineno}

\usepackage[symbol]{footmisc}

\usepackage[none]{hyphenat}
\usepackage{cite}
\usepackage{cleveref}
\usepackage{lineno}

\title{Evidence for a Damped Millisecond Quasi-Periodic Structure in a Fast Radio Burst} 
\author
{Shuo Xiao$^{1,2}$, Zheng-Huo Jiang$^{1,2}$, Di Li$^{3,4,5\ast}$}

\begin{document} 
\begin{CJK}{UTF8}{gbsn}
\sloppy
\baselineskip24pt
\maketitle

\maketitle

\begin{affiliations}
\item School of Physics and Electronic Science, Guizhou Normal University, Guiyang 550001, China
\item Guizhou Provincial Key Laboratory of Radio Astronomy and Data Processing, Guizhou Normal University, Guiyang 550001, China
\item New Cornerstone Science Laboratory, Department of Astronomy, Tsinghua University, Beijing 100084, China
\item National Astronomical Observatories, Chinese Academy of Sciences, Beijing 100101, China
\item Zhejiang Lab, Hangzhou, Zhejiang 311121, China 
\end{affiliations}


\begin{abstract}
Fast radio bursts (FRBs) are millisecond-duration transients of unknown origin, likely associated with compact astrophysical objects. We report evidence of a damped millisecond quasi-periodic structure in the apparently non-repeating FRB~20190122C.
The burst consists of eight closely spaced radio pulses with a characteristic separation of \(\sim3.6\) ms, and the pulse amplitudes show an approximately exponential post-peak decay.
This timing-and-amplitude morphology is broadly consistent with a damped magnetospheric disturbance.
Under an illustrative Alfvénic interpretation, the characteristic timescale corresponds to an order-of-magnitude local magnetic-field estimate of $\sim 10^{12}$ G, broadly compatible with low-field magnetar or strongly magnetized neutron-star scenarios.
The absence of measurable evolution in the pulse spacing and the post-peak damping-like amplitude envelope are less naturally expected in simple merger-driven scenarios. 
These results suggest a rare candidate for an exponentially damped millisecond quasi-periodic structure in an FRB, providing an observational clue to transient oscillatory activity in FRBs.

\end{abstract}
Fast radio bursts (FRBs) are millisecond-duration flashes of radio waves that originate from extragalactic distances. Their immense brightness temperatures imply coherent emission mechanisms, yet the identity of their progenitors remains unresolved. 
Whether all FRBs, especially the repeating and non-repeating types, emerge from a single physical class or represent distinct populations continues to be debated (e.g. \cite{Petroff2019, Cordes2019,zhang2020physical, zhang2023physics}). Repeaters exhibit persistent activity over months or years, whereas some FRBs—despite follow-up efforts—remain singular events. This observational dichotomy suggests a range of possible origins, from cataclysmic explosions to magnetospheric instabilities \cite{platts2019living, feng2022frequency}.

The detection of FRB 200428, temporally coincident with an X-ray burst from the Galactic magnetar SGR J1935+2154, has offered a compelling data point linking magnetars to FRB production \cite{bochenek2020fast, 2020Natur.587...54C, li2021hxmt}. SGR J1935+2154 was subsequently detected as a radio pulsar \cite{zhu2023radio,wang2024x} by the FAST telescope \cite{li2018fast}, the pulses of which can only be seen from within the Milky Way with current technology. Further episodic radio bursts from SGR J1935+2154 have also been reported \cite{kirsten2021sgr1935}.
While this event bridged Galactic and extragalactic phenomena, its radio burst was several orders of magnitude fainter than cosmological FRBs.
A quasi-periodic oscillation at 40 Hz observed in the associated X-ray burst \cite{li2022quasi} raises the possibility that magnetospheric oscillations may manifest across wavebands, though no periodicity was identified in the FRB itself. 

In this work, we present observations of FRB~20190122C (Sec. \ref{Data} in {Methods}), a non-repeating event detected with the CHIME telescope \cite{2024ApJ...969..145C}. 
The burst profile reveals eight narrowly spaced pulses over $\sim$30 milliseconds, with an approximately regular spacing in their arrival times (as shown in \ref{profile}). 
A closer examination shows that the amplitudes decrease approximately exponentially after the strongest, third pulse. This pulse-train morphology is unusual among FRBs and suggests a candidate damped quasi-periodic structure.

To quantify the temporal pattern, we performed peak-timing analyses \cite{chime2022sub,pastor2023fast}. 
{Fitting the Gaussian peak times (Sec. \ref{Pulse modeling} in {Methods}) yields a period of $3.60 \pm 0.05\ \mathrm{ms}$ (as shown in \ref{Fit_TOA} and \ref{tab:pulse_fit_results}). Monte Carlo simulations give a false-alarm probability of $p=0.009$ when conditioned on the observed eight pulse components, and $p=0.01911$ when the number of pulse components is drawn from a Poisson distribution with a mean of eight (as shown in \ref{MC} and \ref{MC_N}; Sec. \ref{Search for periodicity} in {Methods}).
Separately, the post-peak pulse amplitudes are better described by an exponential damping-like envelope with a decay scale of ($\tau_{\rm amp}=2.24\pm0.06$) pulse components (as shown in \ref{Fit_A}, Sec. \ref{Amplitude decay analysis} in {Methods}) at 3.2 $\sigma$ significance (as shown in \ref{MC_A}).
Taken together, these results provide evidence for a candidate damped millisecond quasi-periodic structure in FRB~20190122C, although the timing periodicity alone does not reach a conventional discovery threshold. The nominal timing false-alarm probability is comparable to those reported for several previous intra-burst QPO-like FRB candidates, but should be interpreted cautiously because sample-level trials would further reduce the formal significance. The combination of multiple millisecond-spaced pulses and a clear post-peak damping-like amplitude envelope makes this burst an unusual case among FRBs with fine temporal structure (as shown in \ref{tab:frb_qpo_summary}).}

{Assuming an Alfvén wave interpretation \cite{sotani2008alfven}, the observed frequency implies a magnetic field strength of order $10^{12}$ G (Sec. \ref{Theoretical} in {Methods}), compatible with low-field magnetars such as SGR~0418+5729 \cite{rea2010low}. The characteristic frequency of \(\sim280\) Hz, together with the damping-like amplitude envelope, is reminiscent of transient QPO phenomenology seen in magnetar giant-flare tails, including SGR~1806--20 and SGR~1900+14 \cite{watts2006detection,huppenkothen2014intermittency}. Recent theoretical studies further show that magnetar crustquakes and crust--magnetosphere coupling can excite Alfvénic or fast-magnetosonic disturbances and imprint quasi-periodic signatures on FRB-like emission \cite{qu2025three,burnaz2025crustal}, while FRB propagation in the inner magnetosphere can be attenuated by nonlinear coupling to Alfvénic fluctuations \cite{solanki2026damping}.}
A speculative connection may also be made with the empirical scaling between single-pulse substructure spacing and neutron-star rotation proposed by \cite{2024NatAs...8..230K}. If this scaling is extrapolated to the present millisecond spacing, \(P_{\rm QPO}\approx3.6\) ms would correspond to a possible rotation period of \(\sim3.6\) s, within the range of known magnetars.

Simple binary-merger or compact-object coalescence scenarios appear less natural for this event. Such processes may be expected to show increasing activity as the system approaches merger, whereas FRB~20190122C shows a post-peak damping-like amplitude envelope. Moreover, no measurable evolution in the pulse spacing is seen over the burst duration, which is less naturally expected in rapidly evolving dynamical environments \cite{2016ApJ...822L...7W}.

Our findings suggest that at least some non-repeating FRBs may originate from transient excitations of oscillatory modes in isolated, strongly magnetized neutron stars. 
The observed morphology, approximate regularity, and damping-like amplitude evolution in FRB~20190122C make it a rare case in which transient dynamical processes may be reflected in the radio burst structure.
With only a handful of FRBs exhibiting comparable structure \cite{2021ApJ...919L...6M,chime2022sub, pastor2023fast} (as shown in \ref{guanxi}, \ref{tab:frb_qpo_summary}), events like this may serve as key probes of neutron star interiors and magnetospheres.

\renewcommand{\thefigure}{Figure. \arabic{figure}}
\renewcommand{\figurename}{}
\setcounter{figure}{0}
\begin{figure}
    \centering
    \includegraphics[width=\columnwidth]{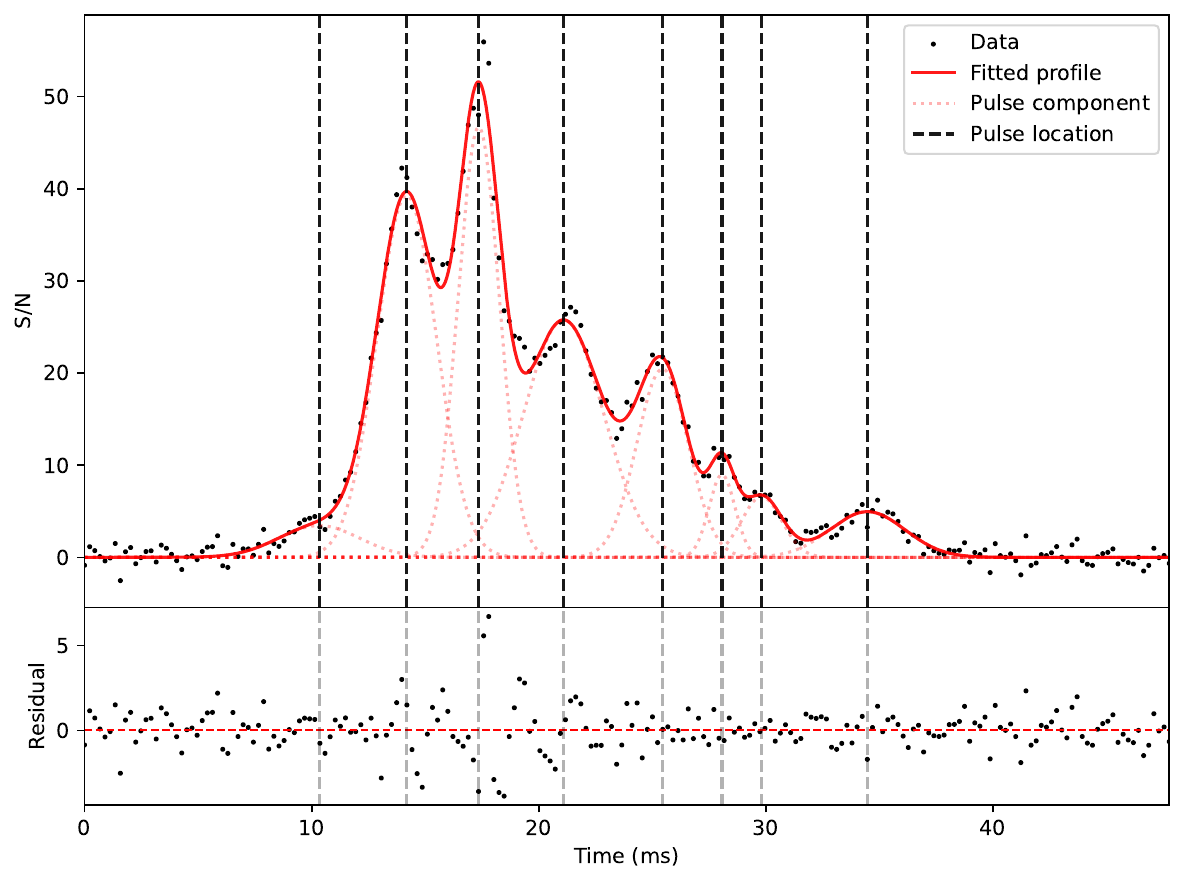}
    \caption{Temporal profile of FRB 20190122C (black points) with Gaussian component fitting (red solid line). The burst is modeled as a sum of multiple Gaussian pulses, each represented by a red-dotted curve. Vertical dashed black lines indicate the location of individual pulse peaks.
}
    \label{profile}
\end{figure}

\begin{figure}
    \centering
    \includegraphics[width=\columnwidth]{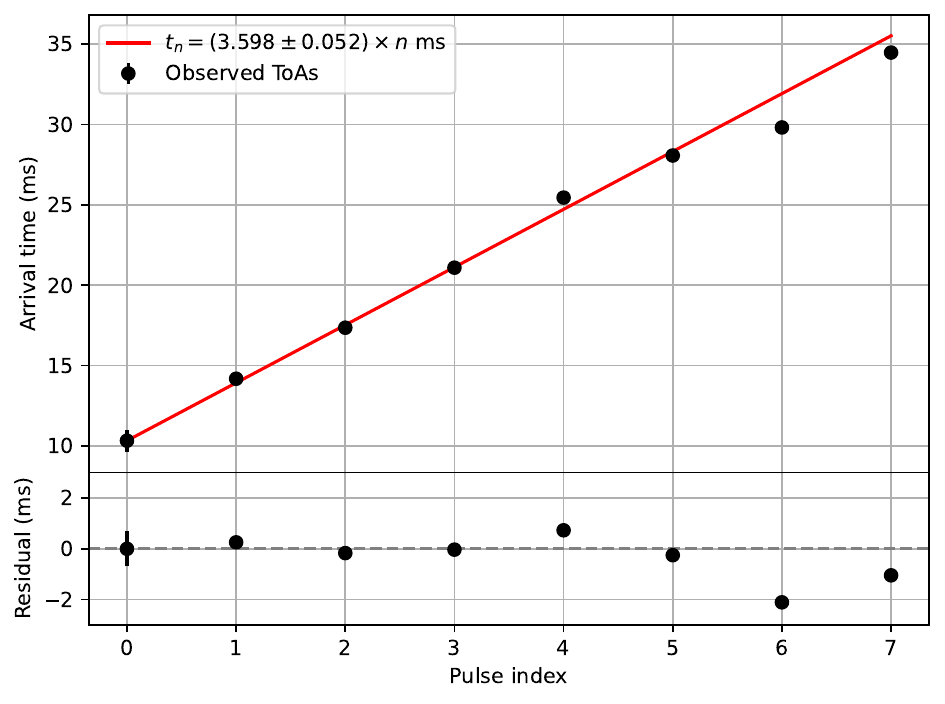}
    \caption{Linear fit to the arrival times of the eight pulse components in the burst. The best-fit model is shown in red with the fitted relation $t_n = T_0 + n P_{\mathrm{QPO}}$, where $P_{\mathrm{QPO}} = ({\text{3.60}} \pm {\text{0.05}})$ ms. The lower panel displays the residuals between the observed and fitted times. }
    \label{Fit_TOA}
\end{figure}

\begin{figure}[http]
\includegraphics[width=\columnwidth]{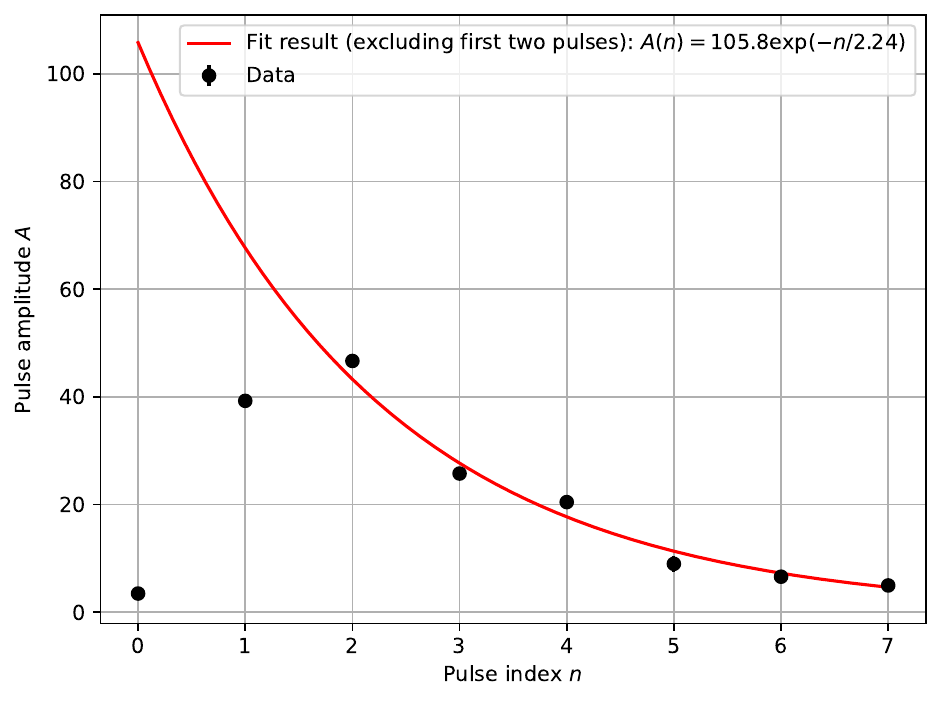}
\caption {Exponential decay of pulse amplitudes. The black circles with error bars denote the measured amplitudes of the eight pulses. The exponential decay fit is applied only to the post-peak components, from the third to the eighth pulse ($n = 2$ to $n = 7$), {because the first two pulses correspond to the rising phase before the burst maximum and may represent the initial triggering or energy-injection stage rather than the subsequent damping phase. The red curve shows the best-fit model $A(n) = A_0 \exp(-n/\tau_{\rm amp})$, with $A_0 = 105.83 \pm 3.80$ and $\tau_{\rm amp} = 2.24 \pm 0.06$ pulse components.}}\label{Fit_A}
\end{figure}

\begin{figure}[http]
\includegraphics[width=\columnwidth]{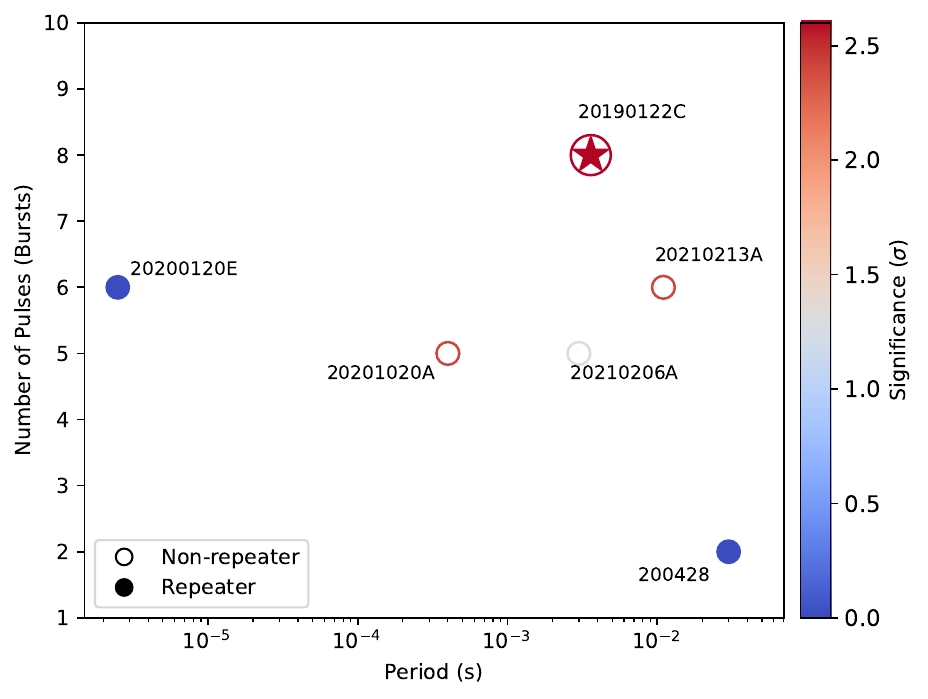}
\caption{Reported intra-burst quasi-periodic or regularly spaced substructures in fast radio bursts \cite{pastor2023fast,2021ApJ...919L...6M,chime2022sub,bochenek2020fast,2020Natur.587...54C}. 
The horizontal axis shows the characteristic period or quasi-period in seconds; the vertical axis indicates the number of resolved pulses or sub-bursts.
Marker colour encodes the reported significance of periodicity (redder implies higher), using a linear colour scale.
Stars denote FRB~20190122C (this work), hollow circles represent non-repeat FRBs, while solid circles mark repeat ones.
Among the comparison sample shown here, FRB~20190122C is the only source for which a millisecond-spaced pulse train is accompanied by a clear post-peak damping-like amplitude envelope.
}\label{guanxi}
\end{figure}

\clearpage
\bibliographystyle{naturemag}

\clearpage
\newpage

\begin{methods}
\section{Data and preprocessing}\label{Data}
FRB~20190122C was detected on 2019 January 22 by the Canadian Hydrogen Intensity Mapping Experiment Fast Radio Burst project (CHIME/FRB). It is localized at R.A. \(=200.4987(12)\) and Decl. \(=17.5914(13)\), with a fluence of \(120(10)~{\rm Jy~ms}\) and a peak flux density of \(13(1)~{\rm Jy}\) \cite{2024ApJ...969..145C}.
Following the public release of baseband-level voltage data\footnote{\url{https://doi.org/10.11570/23.0029}}, we retrieved the coherently dedispersed total intensity time series for analysis. 
{The burst was dedispersed using ${\rm DM}=690.174~{\rm pc~cm^{-3}}$, provided with the CHIME/FRB baseband data product. 
We did not adopt an additional structure-optimising DM as the baseline choice, because this DM already yields a well-resolved burst profile with clearly separated temporal components. 
A test using a structure-optimising DM gives no significant change in the burst profile, component identification, or inferred millisecond spacing.}
These data provide microsecond time resolution and allow detailed investigation of the burst’s fine temporal structure.

Initial processing involved standard radio data handling procedures. We removed frequency channels affected by radio frequency interference (RFI) and integrated across the full observing band to obtain a one-dimensional time series. The dynamic spectrum and the corresponding frequency-integrated burst profile are shown in \ref{waterfall}. The burst emission is visible over a wide frequency range, while the masked frequency channels affected by RFI or poor data quality appear as horizontal blank bands.
This final profile revealed a strikingly structured burst composed of eight sub-pulses, narrowly spaced pulses over a few milliseconds. {The public CHIME/FRB baseband data product does not provide point-by-point uncertainties for the frequency-integrated burst profile. We therefore do not assign artificial error bars to individual time bins, since doing so would require additional assumptions about the noise properties after RFI masking, coherent dedispersion, frequency integration, and normalization. This treatment is consistent with common practice in previous CHIME/FRB burst-profile and FRB quasi-periodicity studies. The timing uncertainties used in our analysis are instead derived from the covariance matrix of the Gaussian pulse-component fits and are propagated into the arrival-time periodicity analysis.}

\section{Check for known pulsar counterparts}\label{sec:pulsar_check}
Recent developments have shown that FRB~20191221A, previously reported as an FRB with prominent periodicity \cite{chime2022sub}, is associated with the known Galactic pulsar PSR~J0248+6021 rather than an extragalactic FRB \cite{frbcollaboration2026series}. This motivates an explicit check for catalogued pulsars near FRB~20190122C.

FRB~20190122C is located at R.A. \(=200.4987(12)\) and Decl. \(=17.5914(13)\). We searched the ATNF pulsar catalogue (psrcat) for all catalogued pulsars within a \(30^\circ\) radius of this position. Among these sources, the pulsar with the highest dispersion measure is PSR~J1239+0326, with \({\rm DM}=34.29~{\rm pc~cm^{-3}}\). By contrast, FRB~20190122C has \({\rm DM}=690.174~{\rm pc~cm^{-3}}\) and lies at a Galactic latitude of \(78.07^\circ\). No catalogued pulsar is therefore consistent with both the localization and dispersion measure of FRB~20190122C.

\section{Pulse modeling}\label{Pulse modeling}
To characterize the temporal morphology, we modeled the burst as a superposition of different Gaussian components. 
Each pulse was fit with the functional form:

\begin{equation}
\begin{split}
f(t; A, \mu, \sigma) = A \exp\left[ -\frac{(t - \mu)^2}{2\sigma^2} \right],
\end{split}
\end{equation}
where $A$ is the amplitude, $\mu$ is the central time (i.e., pulse peak), and $\sigma$ the width. The fitting was performed via nonlinear least-squares optimization, and uncertainties in $\mu$ were estimated from the diagonal terms of the parameter covariance matrix. These peak times, along with their $1\sigma$ uncertainties, form the basis for all time-domain periodicity analyses.

To assess the quality of the fit, we explored the effect of varying the number of pulses included in the model \cite{chime2022sub}. 
{The reduced chi-square values for fits with different numbers of pulse components are shown in \ref{maichong_kafang} and \ref{tab:pulse_fit_results}. The red dashed line at 8 pulses marks the optimal number, where further increases in the number of pulses did not significantly improve the fit quality, indicating that 8 pulses provide a reasonable balance between model complexity and fit accuracy. The adopted eight-component model corresponds to a reduced chi-square of 1.8. Although this value remains somewhat above unity, adding further components does not lead to a substantial improvement in the fit quality. We therefore adopt the eight-component model as a phenomenological description that captures the main pulse structure while avoiding unnecessary model complexity. The corresponding profile residuals are also shown, indicating that the main multi-pulse morphology is reproduced reasonably well, although low-level deviations remain. We also tested an alternative pulse basis using the von Mises function \cite{vonmises1918ganzzahligkeit} and found that the reduced chi-square changes only marginally, from $\chi_\nu^2=1.813$ for the Gaussian basis to $\chi_\nu^2=1.811$ for the von Mises basis. This suggests that the adopted decomposition is phenomenological rather than unique, and that the main pulse locations and the inferred millisecond spacing are not strongly dependent on the specific choice of basis function.}

\section{Search for periodicity}\label{Search for periodicity}
{We assessed the regularity of the eight pulses by fitting their arrival times $t_n$ to a linear model:
\begin{equation}
t_n = T_0 + n P_{\mathrm{QPO}},
\end{equation}
where $T_0$ is the reference arrival time, $n$ is the pulse index (ranging from 0 to 7), and $P_{\mathrm{QPO}}$ is the candidate period. The best-fit period was derived using a weighted least-squares fit, and the fit quality was evaluated via the reduced chi-squared statistic ($\chi^2_\nu$).

To quantify the likelihood of obtaining such regular spacing by chance, we ran Monte Carlo simulations under a null hypothesis in which pulses are spaced randomly but with a minimum delay constraint to account for finite pulse widths. We followed the method proposed by \cite{pastor2023fast}, drawing simulated inter-pulse intervals from a shifted Poissonian distribution:

\begin{equation}
d_i = - (1 - \eta)\,\bar{d} \ln(1 - x_i) + \eta \bar{d},
\end{equation}
where $x_i$ is sampled from a uniform distribution in [0,1), $\bar{d}$ is the mean interval, and $\eta$ controls the minimum spacing (we adopted $\eta = 0.2$ as in \cite{pastor2023fast}). Each synthetic pulse train was subjected to the same linear fitting, and the resulting $\chi^2_\nu$ values were compared against the observed one to estimate a false-alarm probability.

We also applied the same peak-timing search procedure to the publicly available CHIME/FRB baseband sample, which contains 140 bursts. Most of these bursts are single-pulse or two-component events \cite{2024ApJ...969..145C}, or contain too few resolved components for a meaningful intra-burst arrival-time periodicity test. We therefore do not treat all 140 bursts as equivalent trials for this specific analysis. Instead, the effective comparison sample consists of the 10 baseband bursts with more than five identifiable temporal components. Among these multi-component bursts, FRB~20190122C shows the clearest millisecond-spaced timing structure.

As a baseline estimate, 100,000 simulations conditional on the observed number of pulse components, $N=8$, give a nominal false-alarm probability of $p=0.009$ (see \ref{MC}). A more conservative test, in which the pulse number in each simulation is drawn from a Poisson distribution with a mean of eight, gives $p=0.019$ (see \ref{MC_N}). These nominal probabilities are quoted to allow direct comparison with previous FRB intra-burst periodicity studies using similar arrival-time methods. If the 10 multi-component baseband bursts are treated as the effective search sample, the formal significance would be further reduced. We therefore regard the timing regularity as suggestive evidence for a candidate millisecond quasi-periodic substructure, rather than as a robust QPO detection.

Following previous studies of intra-burst periodicity in FRBs, we also tested timing models that allow possible missing cycles between adjacent pulse components. Inserting such gaps does not improve the fit quality or increase the timing significance. This is also true for the interval between the 6th and 7th components, which is shorter than the other separations. We therefore retain the simpler no-gap assignment, in which the shorter 6th--7th interval is treated as part of the timing scatter around a single characteristic spacing.
}

\section{Amplitude decay analysis}\label{Amplitude decay analysis}
{The pulse amplitudes from the Gaussian fits rise during the first two components, reach a maximum at the third component, and then decrease monotonically. We therefore modeled the post-peak amplitude evolution, from the third to the eighth component, using an exponential decay function:

\begin{equation}
A_n = A_0 \exp\left(-\frac{n}{\tau_{\rm amp}}\right),
\end{equation}
where $A_0$ is the amplitude and $\tau_{\rm amp}$ is the decay constant. The first two components were not included in the baseline decay fit, because they correspond to the rising phase before the burst maximum rather than to the subsequent damping phase. Physically, these early components may represent the initial triggering or energy-injection stage, whereas the post-peak components trace the relaxation of the pulse amplitudes. When the first two components are included, the exponential-decay fit is no longer significant, with \(p=0.73\), corresponding to only \(0.34\sigma\), and the fit quality becomes substantially worse. This supports the interpretation that the exponential model describes only the post-peak damping-like envelope, rather than the full amplitude evolution from the initial rise to the decay.

The best-fit value of $\tau_{\rm amp} = 2.24\pm0.06$ pulses was determined via least-squares fitting. To evaluate the significance of the decreasing amplitude envelope, we generated 100,000 mock amplitude sequences drawn from a uniform distribution and applied the same post-peak fitting procedure. Only 0.0016 of simulations produced a better exponential fit, corresponding to a significance of approximately $3.2\sigma$.

We emphasize that this null-hypothesis test does not by itself prove that an exponential decay is the unique possible model. The exponential form was adopted as a physically motivated phenomenological description of a damped amplitude envelope. To test whether other simple decreasing trends can describe the data comparably well, we also fitted the post-peak amplitudes with linear-decay and power-law-decay models. The exponential-decay model gives $\chi^2=71.221$ and $\chi^2_\nu=17.805$. In comparison, the linear-decay model gives $\chi^2=286.580$ and $\chi^2_\nu=71.645$, while the power-law-decay model gives $\chi^2=114.193$ and $\chi^2_\nu=28.548$. Thus, among the simple monotonic decay models tested, the exponential model provides the best phenomenological description of the post-peak amplitude evolution.}

We note that the similar fast-rise and exponential-decay morphologies have also been recently reported in a Galactic radio burst in the Carina nebula \cite{rajwade2025carina}, suggesting that such temporal envelopes may be a more generic signature of magnetospheric relaxation processes.

\section{Theoretical interpretation and parameter estimates}\label{Theoretical}
{Given the observed quasi-periodic spacing $P_{\rm QPO} \approx 3.60$ ms and the approximately exponential post-peak amplitude decay, the pulse train is consistent with a short-lived, damped disturbance in a neutron-star crust or magnetosphere, as expected in magnetar models involving Alfvén-wave propagation, crustal shear oscillations, and magneto-elastic coupling \cite{Levin2007,van2012magnetar,sotani2008alfven}. A localized crustal or magnetospheric energy release may excite elastic, Alfvénic, or fast-magnetosonic perturbations, which can modulate the plasma conditions, particle acceleration, charge bunching, or the coherent-emission region. Recent simulations further show that magnetar crustquakes and crust--magnetosphere coupling can transmit Alfvénic disturbances into the magnetosphere and imprint quasi-periodic signatures on FRB-like signals \cite{qu2025three,burnaz2025crustal}. In addition, FRB propagation through the inner magnetosphere may be attenuated through nonlinear coupling to Alfvénic fluctuations \cite{solanki2026damping}. These studies provide a possible physical link between damped magnetar oscillations and coherent radio emission, although the present data do not uniquely determine the emission mechanism.}

If the candidate quasi-periodic spacing is associated with standing Alfvénic modes in a magnetar magnetosphere (e.g. \cite{sotani2008alfven}), we can estimate the corresponding local magnetic-field scale. The Alfvén velocity is given by $v_A = B/\sqrt{4\pi \rho}$, where $\rho$ is the plasma mass density. For a loop or field-line structure of scale $L$, the fundamental oscillation frequency is $f_{\mathrm{QPO}} = v_A / 2L$, which leads to
\begin{equation}
B = 2 L f_{\mathrm{QPO}} \sqrt{4\pi \rho}.
\end{equation}

Adopting representative values of $L \sim 10^6$~cm and $\rho \sim 10^5$~g\,cm$^{-3}$ and using the observed QPO frequency $f_{\mathrm{QPO}} = 1/P_{\mathrm{QPO}} \approx 3\times 10^2$~Hz, we obtain
\begin{equation}
B \approx 2 \times 10^6~\mathrm{cm} \times 3 \times 10^2~\mathrm{Hz} \times \sqrt{4\pi \times 10^5~\mathrm{g\,cm^{-3}}} \approx 10^{12}~\mathrm{G}.
\end{equation}

The resulting field estimate, \(B\sim10^{12}~{\rm G}\), is lower than the canonical dipole fields of many Galactic magnetars (\(10^{14}\)--\(10^{15}~{\rm G}\)). However, this value should be interpreted as a model-dependent local or effective field associated with the assumed oscillating region, rather than as a direct measurement of the global dipole field. Some magnetar candidates, such as SGR~0418+5729, have inferred dipole fields as low as \(\sim8\times10^{12}~{\rm G}\) while still displaying magnetar-like activity \cite{rea2010low}. Moreover, the dipole field inferred from spin-down is not necessarily the sole reservoir of magnetic energy; strong internal toroidal components and twisted magnetospheres may store substantially more energy than indicated by the external dipole alone. Recent theoretical work has suggested that intermediate-field magnetars (\(B\sim10^{12-13}~{\rm G}\)) may still power energetic bursts through processes such as magnetospheric reconnection, Alfvén--fast-mode conversion, or magneto-elastic oscillations in the stellar crust \cite{wadiasingh2020frb,wang2023intermediate}.
In such scenarios, quasi-periodic pulse trains could arise naturally from oscillatory modes or reconnection-driven plasmoid shedding, while the observed exponential damping reflects dissipation of these transient modes. 
Thus, even if the surface dipole field of FRB 20190122C is relatively modest, the event remains consistent with a magnetar origin where the key driver is internal magnetic structure and magnetospheric dynamics, rather than the dipole strength alone.

This magnetic-field estimate is sensitive to the assumed oscillation scale \(L\) and local plasma density \(\rho\), since \(B\propto L\rho^{1/2}\). A denser plasma or a larger effective oscillation scale would increase the inferred field, whereas a more compact oscillation region would reduce it. Furthermore, if the observed spacing corresponds to a higher-order mode rather than the fundamental, the inferred field strength would change by the corresponding mode factor. The field estimate should therefore be regarded only as an order-of-magnitude indication rather than a direct measurement of the global dipole field. 
Taken together, the inferred parameters remain broadly consistent with Alfvénic disturbances in a strongly magnetized neutron-star magnetosphere, and therefore support a magnetar-like interpretation as one plausible scenario for FRB~20190122C.

This finding is reminiscent of oscillatory behavior observed in magnetar X-ray bursts, where quasi-periodic signals have been interpreted as global seismic or magnetospheric oscillations. If the candidate radio quasi-periodic structure originates from Alfvénic oscillations in a magnetized plasma, the implied field scale would be broadly compatible with magnetar-like environments, lending support to the possibility that some apparently non-repeating FRBs may originate from energetic activity in isolated compact objects.

An intriguing possibility arises when considering the quasi-periodic pulse spacing in FRB~20190122C in the context of neutron star rotation. \cite{2024NatAs...8..230K} proposed that sub-millisecond quasi-periodic features in single pulses may scale with the spin period of neutron stars via the empirical relation \( P_{\mu} \approx 10^{-3} P_{\mathrm{rot}} \), where \( P_{\mu} \) is the substructure spacing and \( P_{\mathrm{rot}} \) is the rotation period. Applying this scaling to our detected \( P_{\mathrm{QPO}} \approx 3.6\,\mathrm{ms} \), we infer a potential spin period of \( P_{\mathrm{rot}} \approx 3.6\,\mathrm{s} \). This lies within the range of known radio pulsars and low-field magnetars such as SGR~0418+5729.

Another plausible class of FRB progenitor models involves compact binary mergers or interactions, such as neutron star–neutron star coalescence or magnetospheric collisions during inspiral phases (e.g. \cite{2013PASJ...65L..12T, 2016ApJ...822L...7W}). However, these scenarios appear less natural for FRB~20190122C. First, the candidate quasi-periodic spacing shows no measurable evolution over the $\sim$30 ms duration of the burst, which is less naturally expected for a rapidly evolving dynamical timescale in merger-driven processes.
Second, if the observed pulse train originated from magnetospheric interactions between inspiraling compact objects, the energy release would likely increase toward coalescence, resulting in a rising amplitude envelope. 
In contrast, the pulse amplitudes in FRB 20190122C show an exponential decay following the strongest pulse, consistent with a relaxation or damping process rather than escalating activity. 
These features are more naturally explained by a transient excitation of quasi-normal modes in an isolated neutron star magnetosphere, rather than merger-driven dynamics.

\section{Data availability}
The full-resolution baseband data and dedispersed time series of FRB~20190122C analyzed in this study are publicly released by the CHIME/FRB Collaboration \url{https://doi.org/10.11570/23.0029}. No proprietary data were used.

\section{Code availability}
All codes developed for data analysis and figure generation are available from the corresponding author upon reasonable request. The method for evaluating QPO significance follows the Monte Carlo approach introduced in \cite{pastor2023fast}, with modifications tailored to the pulse train structure of FRB~20190122C.

\end{methods}


\begin{addendum}
\item We acknowledge the public data from CHIME. This work is supported by the National Natural Science Foundation of China (NOs. 12588202, 12573043, 12303043), Science and Technology Foundation of Guizhou Province (Key Program, No. ZK[2024]430), the Natural Science Research Project of the Guizhou Provincial Department of Education ([2024]321). 

\item[Correspondence] Correspondence and requests for materials should be addressed to (D. L,, email: dili@tsinghua.edu.cn).

\item[Author Contributions]  D. L. initiated this project. S. X. led the data analysis and statistical analysis. D. L. led the organization of the manuscript and theory interpretation.
All authors participated in the discussion and contributed to the data analysis which is important for this work.

\item[Competing Interests] The authors declare no competing interests.

\end{addendum}


\newpage

\renewcommand{\thefigure}{Extended Data Figure. \arabic{figure}}
\renewcommand{\figurename}{}
\setcounter{figure}{0}

\begin{figure}
    \centering
    \includegraphics[width=\columnwidth]{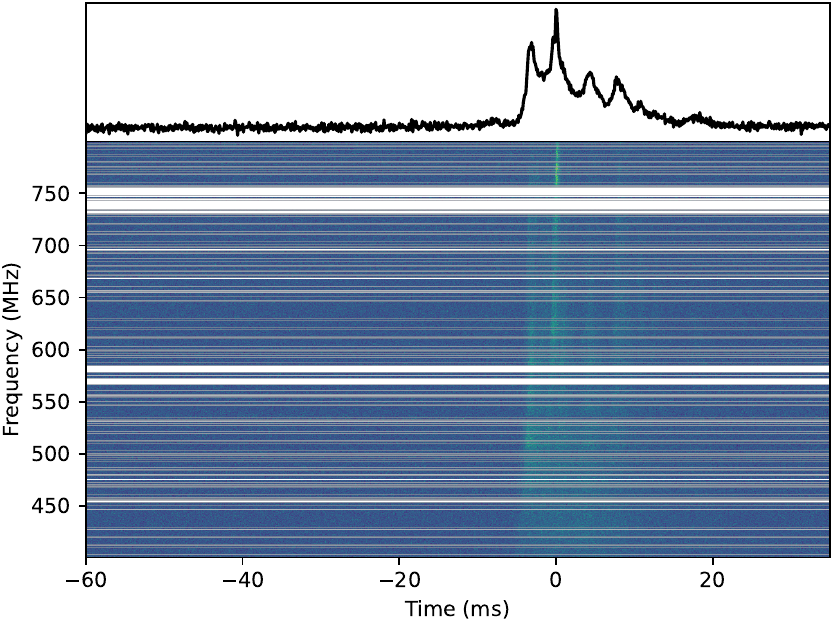}
    \caption{Dynamic spectrum of FRB 20190122C. The lower panel shows the frequency–time intensity map, and the upper panel shows the frequency-integrated burst profile. The burst is visible over a wide frequency range, and white horizontal bands denote masked frequency channels affected by radio-frequency interference or poor data quality.
}
    \label{waterfall}
\end{figure}

\begin{figure}
    \centering
    \includegraphics[width=\columnwidth]{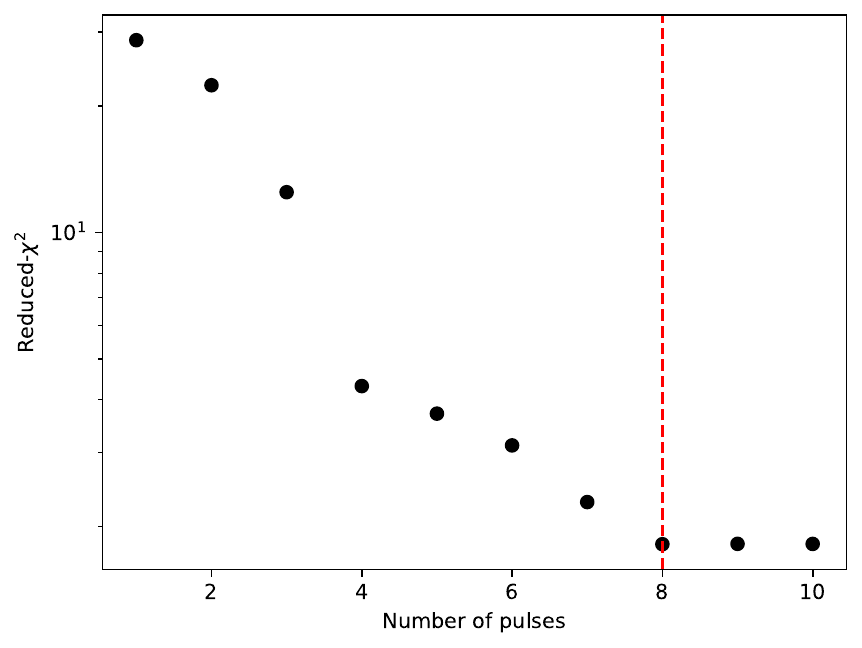}
    \caption{Reduced-\(\chi^2\) values for fits with different numbers of pulses. The x-axis represents the number of pulses used in the fit, and the y-axis shows the corresponding reduced \(\chi^2\) values on a logarithmic scale. The red dashed vertical line at 8 pulses marks the optimal number of pulses, where further increases in the number of pulses do not significantly improve the fit quality. This suggests that 8 pulses provide a balance between model complexity and fit accuracy.}
    \label{maichong_kafang}
\end{figure}

\begin{figure}
    \centering
    \includegraphics[width=\columnwidth]{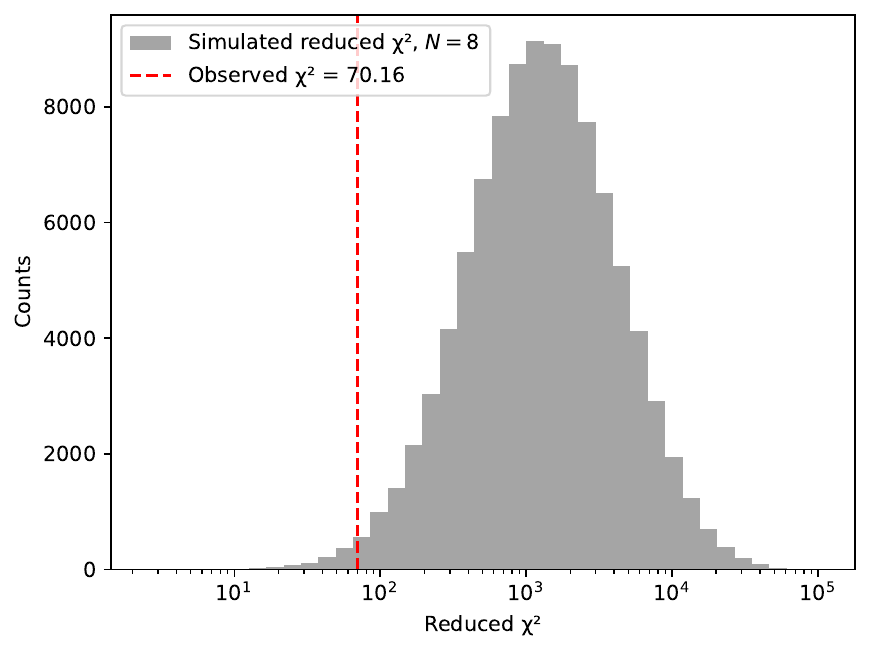}
    \caption{Monte Carlo simulation of the reduced $\chi^2$ statistic for the arrival-time regularity test. The histogram shows the distribution of reduced $\chi^2$ values from 100,000 Monte Carlo simulations under the null hypothesis of randomly spaced pulses, conditional on the same number of pulse components as observed in FRB 20190122C, $N$ = 8. The observed $\chi^2_\nu$ is marked with a red dashed line and lies in the tail of the distribution, corresponding to a false-alarm probability of \(p=0.009\).}
    \label{MC}
\end{figure}

\begin{figure}
    \centering
    \includegraphics[width=\columnwidth]{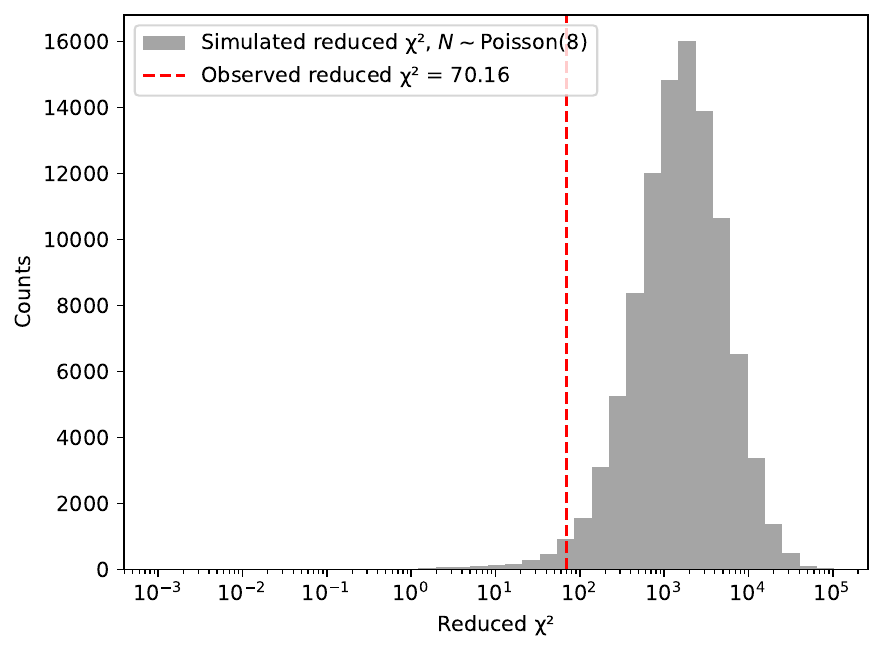}
    \caption{Monte Carlo simulation of the reduced $\chi^2$ statistic for the arrival-time regularity test without fixing \(N\). The histogram shows the distribution of reduced $\chi^2$ values from 100,000 Monte Carlo simulations under the null hypothesis of randomly spaced pulses. The total number of pulse components in each simulation was drawn from a Poisson distribution with a mean of eight, matching the observed number of components in FRB 20190122C. The simulated pulse arrival times were analysed using the same linear timing-fit procedure as applied to the observed data. The observed $\chi^2_\nu$ is marked with a red dashed line and lies in the tail of the simulated distribution, corresponding to a false-alarm probability of $p=0.019$.}
    \label{MC_N}
\end{figure}


\begin{figure}
    \centering
    \includegraphics[width=\columnwidth]{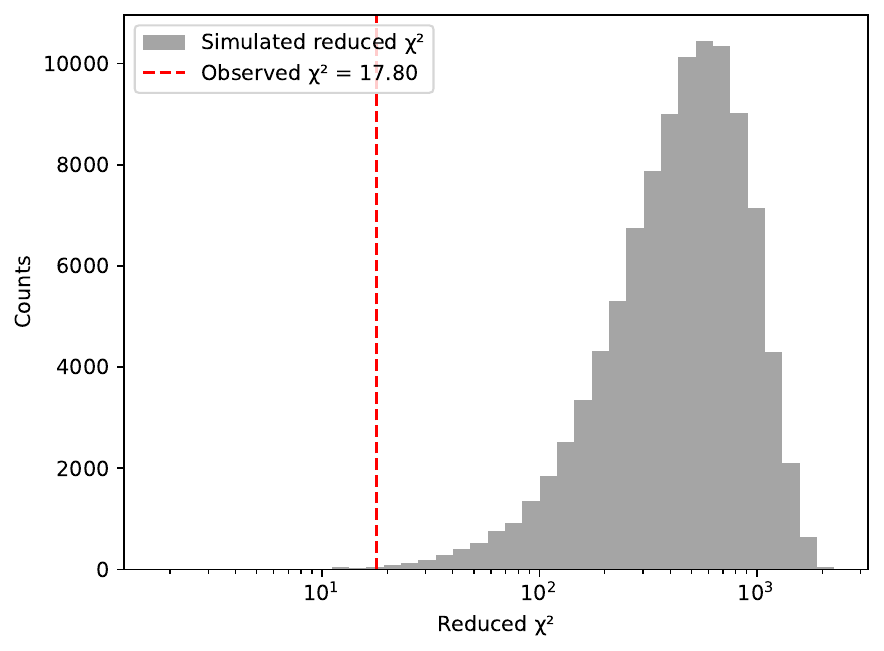}
    \caption{Monte Carlo simulation of the reduced $\chi^2$ statistic for exponential amplitude decay. The histogram shows the distribution of reduced $\chi^2$ values obtained by fitting exponential models to 100,000 simulated pulse amplitude sequences drawn from a uniform distribution. These simulations follow the null hypothesis of randomly fluctuating pulse strengths without any systematic decay. The vertical red dashed line marks the reduced $\chi^2$ value measured from the observed data, which lies in the extreme tail of the distribution. This corresponds to a false-alarm probability of $p \approx 1.6 \times 10^{-3}$.}
    \label{MC_A}
\end{figure}
\renewcommand{\thetable}{Extended Data Table. \arabic{table}}
\renewcommand{\tablename}{}
\setcounter{table}{0}
\clearpage

\begin{table*}[tb]
\centering
\scriptsize
\setlength{\tabcolsep}{2pt}
\renewcommand{\arraystretch}{2}
\renewcommand{\tabularxcolumn}[1]{m{#1}}
\caption{Best-fit pulse amplitudes, pulse centroids, reduced $\chi^2$ values of the burst-profile fitting, and corresponding quasi-periodicity $p$-values for decompositions with different assumed numbers of pulse components. For each model, the fitted amplitudes $A$ and pulse centroids $\mu$ are listed together with the reduced $\chi^2$ from the light-curve profile fit and the $p$-value from the timing analysis. This comparison is intended to show how both the profile-fit quality and the formal significance of the candidate quasi-periodic structure vary with the adopted phenomenological decomposition of the burst profile.}
\label{tab:pulse_fit_results}
\begin{tabularx}{\textwidth}{
>{\centering\arraybackslash}m{1.1cm}
>{\centering\arraybackslash}X
>{\centering\arraybackslash}X
>{\centering\arraybackslash}m{1.3cm}
>{\centering\arraybackslash}m{1.1cm}
}

\toprule
Pulse number & Amplitude $A$ & Mean $\mu$ (ms) & Reduced $\chi^2$ & Quasi-periodicity $p$ value \\
\midrule

3 &
\makecell[c]{$39.61 \pm 0.55$, $18.28 \pm 0.27$, $4.82 \pm 0.38$} &
\makecell[c]{$15.93 \pm 0.06$, $23.98 \pm 0.17$, $34.81 \pm 0.12$} &
12.48 &
0.308 \\

4 &
\makecell[c]{$1.03 \pm 0.88$, $28.92 \pm 0.57$, $33.91 \pm 0.58$, \\
$22.04 \pm 0.22$} &
\makecell[c]{$9.72 \pm 0.25$, $14.10 \pm 0.02$, $17.32 \pm 0.02$,\\
$21.02 \pm 0.11$} &
4.31 &
0.183 \\

5 &
\makecell[c]{$1.29 \pm 0.84$, $30.11 \pm 0.59$, $35.40 \pm 0.60$,\\
$7.25 \pm 0.65$, $19.99 \pm 0.33$} &
\makecell[c]{$9.73 \pm 0.21$, $14.09 \pm 0.02$, $17.33 \pm 0.02$,\\
$21.31 \pm 0.05$, $21.35 \pm 0.14$} &
3.71 &
0.476 \\

6 &
\makecell[c]{$1.83 \pm 0.74$, $30.54 \pm 0.66$, $35.34 \pm 0.63$, \\
$6.78 \pm 0.67$, $20.58 \pm 0.32$, $4.21 \pm 0.43$} &
\makecell[c]{$9.69 \pm 0.17$, $14.09 \pm 0.02$, $17.34 \pm 0.02$, \\
$21.33 \pm 0.05$, $21.34 \pm 0.15$, $34.97 \pm 0.12$} &
3.12 &
0.289 \\

7 &
\makecell[c]{$38.60 \pm 0.37$, $43.73 \pm 1.00$, $25.60 \pm 0.39$, \\
$20.08 \pm 0.55$, $9.18 \pm 1.36$, $6.55 \pm 0.56$,\\
$4.95 \pm 0.35$} &
\makecell[c]{$14.22 \pm 0.03$, $17.40 \pm 0.02$, $21.04 \pm 0.05$,\\
$25.47 \pm 0.05$, $28.06 \pm 0.10$, $29.83 \pm 0.23$, \\
$34.47 \pm 0.15$} &
2.28 &
0.047 \\

8 &
\makecell[c]{$3.45 \pm 0.44$, $39.27 \pm 0.71$, $46.67 \pm 0.94$, \\
$25.75 \pm 0.40$, $20.44 \pm 0.51$, $8.97 \pm 1.43$, \\
$6.58 \pm 0.55$, $4.96 \pm 0.35$
} &
\makecell[c]{$10.33 \pm 0.69$, $14.18 \pm 0.04$, $17.35 \pm 0.02$, \\
$21.09 \pm 0.05$, $25.45 \pm 0.05$, $28.07 \pm 0.10$, \\
$29.81 \pm 0.23$, $34.47 \pm 0.15$
} &
1.81 &
0.009 \\

9 &
\makecell[c]{$3.43 \pm 0.43$, $39.29 \pm 0.65$, $46.42 \pm 0.96$, \\
$25.74 \pm 0.40$, $2.55 \pm 1.29$, $20.37 \pm 0.53$, \\
$9.61 \pm 1.26$, $6.48 \pm 0.60$, $4.94 \pm 0.35$
} &
\makecell[c]{$10.29 \pm 0.66$, $14.18 \pm 0.04$, $17.35 \pm 0.02$, \\
$21.11 \pm 0.05$, $24.10 \pm 0.11$, $25.50 \pm 0.06$, \\
$28.06 \pm 0.11$, $29.87 \pm 0.23$, $34.46 \pm 0.15$
} &
1.82 &
0.153 \\

10 &
\makecell[c]{$3.43 \pm 0.43$, $39.30 \pm 0.64$, $46.39 \pm 0.96$, \\
$25.73 \pm 0.40$, $2.60 \pm 1.31$, $20.36 \pm 0.53$, \\
$9.87 \pm 1.08$, $6.61 \pm 0.59$, $2.07 \pm 0.79$, \\
$5.20 \pm 0.39$
} &
\makecell[c]{$10.29 \pm 0.65$, $14.18 \pm 0.04$, $17.35 \pm 0.02$, \\
$21.11 \pm 0.05$, $24.10 \pm 0.10$, $25.50 \pm 0.06$, \\
$28.08 \pm 0.11$, $29.95 \pm 0.21$, $32.18 \pm 0.19$, \\
$34.68 \pm 0.15$
} &
1.82 &
0.284 \\

\bottomrule
\end{tabularx}
\end{table*}


\begin{table}
\centering
\caption{Representative FRBs with reported intra-burst quasi-periodic or regularly spaced substructures. For FRB~200428, the quoted timescale refers to the separation between two sub-pulses rather than a true period.}
\begin{adjustbox}{width=\columnwidth,center}
\begin{tabular}{ccccccc}
\hline
FRB Name & Period & Pulses & QPO significance ($\sigma$) &Exp. Decay significance ($\sigma$) &  Repeater & Citation \\
\hline
FRB~20190122C &3.6 ms     & 8     & 2.6 & 3.2     & No   & this work \\
FRB~20210206A    &3 ms &5 &1.3&None &No & \cite{chime2022sub}\\
FRB~20210213A    &11 ms &6 &2.4&None  &No & \cite{chime2022sub}\\
FRB~20201020A         & 0.4 ms    & 5     & 2.5&None    & No   &  \cite{pastor2023fast}\\
FRB~20200120E         & $2-3~\mu$s  & 6  & None &None    & Yes  & \cite{2021ApJ...919L...6M} \\
FRB~200428 & 30 ms & 2 & None&None &  Yes & \cite{bochenek2020fast,2020Natur.587...54C}\\
\hline
\end{tabular}
\end{adjustbox}
\label{tab:frb_qpo_summary}
\end{table}

\end{CJK}
\end{document}